\documentclass[%
 reprint,
superscriptaddress,
 amsmath,amssymb,
pra,
]{revtex4-2}
\usepackage{url}
\usepackage{braket}
\usepackage{graphicx}
\usepackage{dcolumn}
\usepackage{bm}
\usepackage{color,soul}
\usepackage{hyperref}
\hypersetup{
    colorlinks=true,
    linkcolor=blue,
    filecolor=magenta,      
    urlcolor=cyan,
}

\begin{document}

\preprint{APS/123-QED}

\title{Single-Hole Pump in Germanium}

\author{Alessandro Rossi}
\email{alessandro.rossi@strath.ac.uk}
\affiliation{
Department of Physics, SUPA, University of Strathclyde, Glasgow G4 0NG, United Kingdom}
\affiliation{National Physical Laboratory, Hampton Road, Teddington TW11 0LW, United Kingdom
}
\author{Nico W. Hendrickx}%
\affiliation{
QuTech and Kavli Institute of Nanoscience, Delft University of Technology, Delft, The Netherlands
}
 \author{Amir Sammak}
 \affiliation{
QuTech and Netherlands Organisation for Applied Scientific Research (TNO), Delft, The Netherlands
}
 \author{Menno Veldhorst}
 \affiliation{
QuTech and Kavli Institute of Nanoscience, Delft University of Technology, Delft, The Netherlands
}
 \author{Giordano Scappucci}
 \affiliation{
QuTech and Kavli Institute of Nanoscience, Delft University of Technology, Delft, The Netherlands
}
 \author{Masaya Kataoka}%
\affiliation{National Physical Laboratory, Hampton Road, Teddington TW11 0LW, United Kingdom
}%
\begin{abstract}
Single-charge pumps are the main candidates for quantum-based standards of the unit ampere because they can generate accurate and quantized electric currents. In order to approach the metrological requirements in terms of both accuracy and speed of operation, in the past decade there has been a focus on semiconductor-based devices. The use of a variety of semiconductor materials enables the universality of charge pump devices to be tested, a highly desirable demonstration for metrology, with GaAs and Si pumps at the forefront of these tests. Here, we show that pumping can be achieved in a yet unexplored semiconductor, i.e. germanium. We realise a single-hole pump with a tunable-barrier quantum dot electrostatically defined at a Ge/SiGe heterostructure interface. We observe quantized current plateaux by driving the system with a single sinusoidal drive up to a frequency of 100 MHz. The operation of the prototype was affected by  accidental formation of multiple dots, probably due to disorder potential, and random charge fluctuations. We suggest straightforward refinements of the fabrication process to improve pump characteristics in future experiments.
\end{abstract}

\maketitle


\section{\label{sec:Intro}Introduction}
A single-charge pump is an electronic device that can generate quantized electric current by clocking the transfer of individual charged quasi-particles (electrons, holes or cooper pairs) with an external periodic drive~\cite{pekola-rev}. The pumped current can be expressed as $I=nef$, where $e$ is the elementary charge, $f$ is the frequency of the drive and $n$ is an integer representing  the number of particles transferred per period. The development of this technology has been mainly motivated by its possible application for quantum-based standards of electric current~\cite{scherer-rev}. To date, the most promising pump realisations are semiconductor quantum dots (QDs) with tunable tunnel barriers~\cite{zhao,Giblin_2019,gento-acc,stein-2015}, which have demonstrated to operate at the highest frequencies (up to few GHz) with the lowest current uncertainty (below part per million), in the pursuit of meeting the stringent metrological requirements~\cite{Scherer_2012}.\\\indent
At the core of the any quantum standard lies the concept of universality. This is the idea that the operation of the standard is based on fundamental principles of nature, rather than being dependent on its specific physical implementation. For example, the acceptance of the Quantum Hall Effect as a primary standard of resistance was driven by the experimental demonstrations of agreement at a level of relative uncertainty below $10^{-10}$ among Hall devices manufactured in silicon and GaAs, and later in graphene~\cite{Si-hall,Janssen_2012}. Similarly, universality expects that the quantised currents generated by singe-charge pumps do not depend on the material system used. Recently, a study of this kind has shown that there is agreement at a level of $\approx10^{-6}$ between silicon and GaAs pumps~\cite{Giblin_2019}, an encouraging stepping stone for future tests with higher accuracy. Hence, from a universality standpoint, it is of interest to investigate a range of material and device systems that can support clocked charge transfers. In fact, besides the advanced performance achieved with QD pumps, there have been other less fulfilling demonstrations in superconductors~\cite{cooper1,cooper2}, normal metal~\cite{Pothier_1992}, hybrid normal/superconductive metal~\cite{hybrid}, single atoms~\cite{Tettamanzi_2014,roche}, and graphene~\cite{graphene}.\\\indent
Here, we introduce a new material system to the family of single-charge pumps, i.e. germanium (Ge). We demonstrate single-hole transfers clocked by a single sinusoidal drive in a tunable-barrier QD electrostatically formed at a Ge/SiGe heterostructure interface. We ascertain that the value of the current plateaux scales linearly with the rf drive frequency up to $100$~MHz, as expected for quantized transport. We observe unusual plateaux boundaries in the 2D pump maps and tentatively attribute them to multiple parallel pump operation. We also observe device instability due to random charge fluctuations and suggest changes in the fabrication of the next generation of pumps that may mitigate this problem.
\begin{figure*}[t]
\includegraphics[scale=1.1]{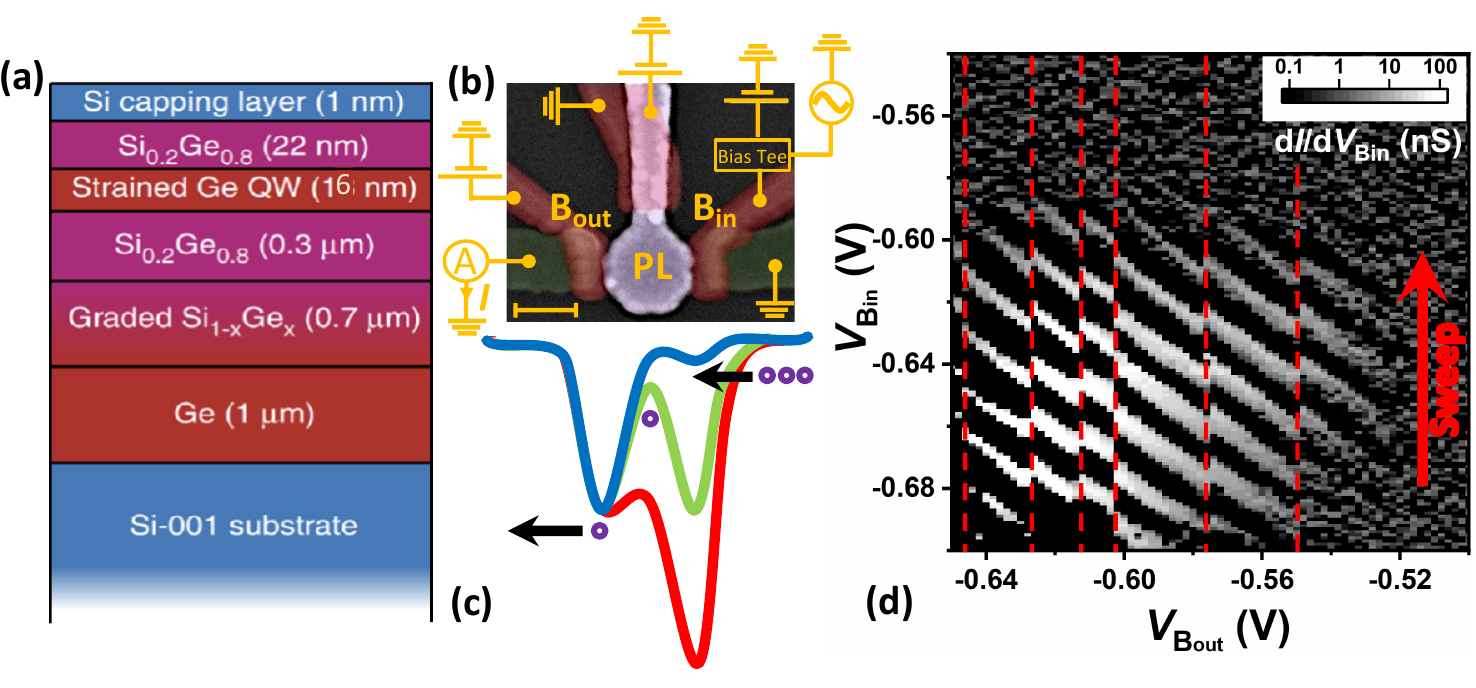}
\caption{(a) Schematic depiction of the Ge/SiGe heterostructure stack. (b)  SEM image of the device gate layout and schematic of the experimental set-up. Different metallization layers are color coded: Al (green), Ti/Pd layer 1 (red), Ti/Pd layer 2 (magenta). Scale bar represents 100 nm.  (c) Valence band energy profile during a pumping cycle: loading (blue), trapping (green), ejection (red). Holes are represented by empty circles. (d) Device transconductance as a function of $V_\textup{Bout}$ and $V_\textup{Bin}$ for $V_\textup{PL}=-1.2$~V and $V_\textup{bias}=1$~mV. Red dashed lines are guides to the eye to highlight random switching events. The vertical arrow shows the direction of the voltage sweeps.}
\label{fig:dc}
\end{figure*}
\section{\label{sec:Methods}Methods}
The sample used in the experiments was fabricated on a Ge/SiGe heterostructure grown on a 100-mm-thick n-type Si(001) substrate~\cite{amir2019}. The material stack is composed of a 16-nm-thick compressively strained Ge quantum well grown on a strain-relaxed Si$_{0.2}$Ge$_{0.8}$ buffer layer. The Ge quantum well (QW) is undoped and separated from the semiconductor/oxide interface by a 22~nm Si$_{0.2}$Ge$_{0.8}$ barrier, as shown in Fig.~\ref{fig:dc}(a). A two-dimensional hole gas with densities up to $6\times10^{11}$ cm$^{-2}$, transport mobility up to $5\times10^5$ cm$^2$/Vs, and effective mass of $\approx 0.05~m_e$, where $m_e$ is the free electron mass, is accumulated in the Ge QW via electrostatic gating~\cite{lodari2019}. The device's gate layout is shown in Fig.~\ref{fig:dc}(b). Ohmic contacts (shown in green) are defined by electron beam lithography, electron beam evaporation and lift-off of a 30-nm-thick Al layer. Electrostatic gates consist of two Ti/Pd layers with thickness of 20~nm and 40~nm for the barrier (red) and plunger gate (magenta) layer, respectively. Both layers are separated from the substrate and each other by 10~nm of ALD-grown Al$_2$O$_3$~\cite{nico2020}.\\ \indent
The measurement set-up is schematically represented in Fig.~\ref{fig:dc}(b). The metal gates are connected to programmable dc voltage sources through
room-temperature low-pass filters (not shown). The gate voltages are used to selectively accumulate holes in the Ge QW, resulting in the formation of tunnel barriers (under gates B\textsubscript{in} and B\textsubscript{out}) that separate a quantum dot (under PL) from the hole reservoirs. Note that one remaining gate is kept at ground potential at all times to laterally confine the QD in the orthogonal direction to transport. The device current, $I$, is measured by a low-noise transimpedance amplifier connected to an ohmic contact. Gate B\textsubscript{in} is connected to an rf source through a low-temperature bias-tee. This gate operates as the entrance barrier for the pumping cycle by clocking the loading of holes into the QD. The pumping protocol used in this work is known as the ratchet mode, which typically applies to single-QD single-drive pumps and is largely insensitive to device bias~\cite{Kaestner_2015}. Each pumping cycle begins with the rf drive rising the potential of the entrance barrier loading the QD with holes from the nearest reservoir. Then the rf drive lowers the barrier to trap holes and eject them across the exit barrier, B\textsubscript{out}, and onto the other reservoir, as depicted in Fig.~\ref{fig:dc}(c). Unless otherwise stated, the measurements presented in this work are carried out with only a small stray bias across device ohmics due to the amplifier ($V_\textup{bias}\approx 250~\mu$V), no intended bias voltage is otherwise applied during pumping. The sample is cooled in a cryogen-free dilution refrigerator with a base temperature of approximately 12 mK, although the effective device temperature may be higher due to heating generated by the rf drive.
\section{\label{sec:Results}Results}
\begin{figure}[t]
\includegraphics[scale=0.56]{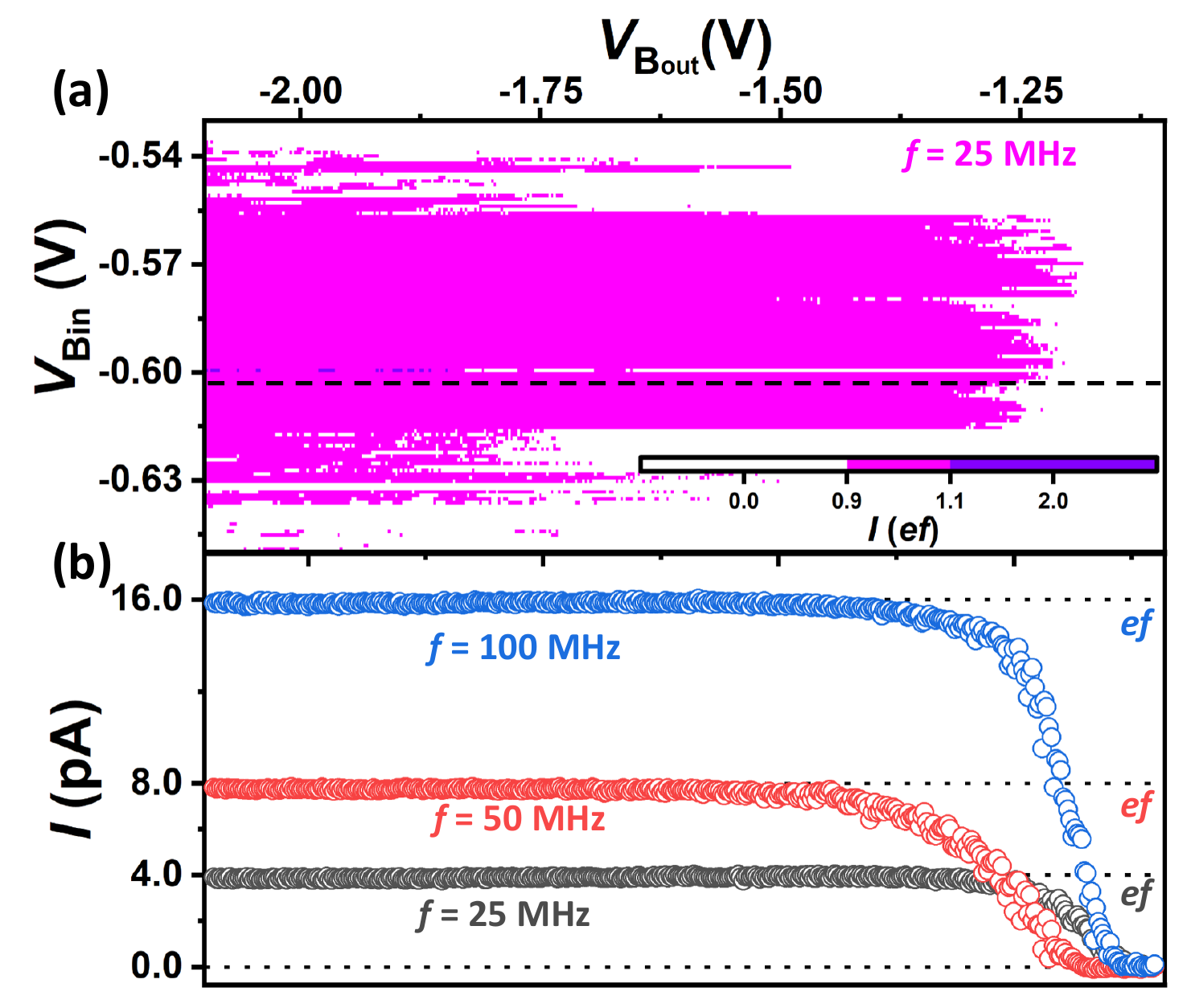}
\caption{(a) Pumped current as a function of entrance and exit barrier gate voltages at $f = 25$~MHz. Black dashed line is a guide for the eye to indicate the region of parameter space of panel (b). The other experimental parameters are $V_\textup{PL}=-1.40$~V, $V_\textup{pp}=0.275$~V. (b) Pumped current as a function of $V_\textup{Bout}$ at $f = 25$~MHz (black circles), $f = 50$~MHz (red), $f = 100$~MHz (blue). Other experimental parameters are $V_\textup{pp}=0.275$~V and $V_\textup{Bin}=-0.602$~V. Black dotted lines indicate the current values expected for quantized pumping, $I=ef$, at the drive frequencies used in the experiment.}
\label{fig:freq}
\end{figure}

\begin{figure}[t]
\includegraphics[scale=0.66]{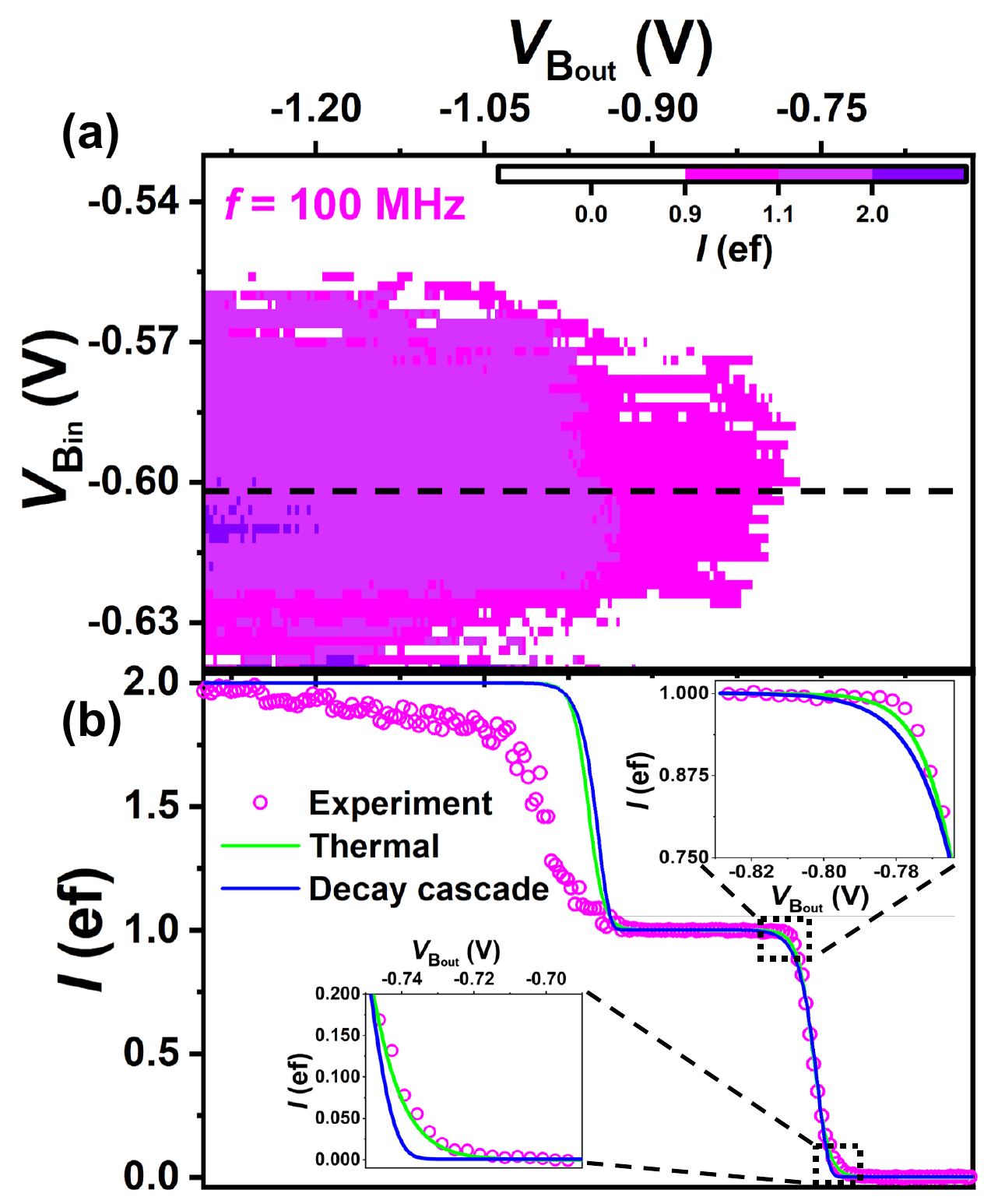}
\caption{(a) Pumped current as a function of entrance and exit barrier gate voltages at $f=100$~MHz. Black dashed line is a guide for the eye to indicate the region of parameter space of panel (b). The other experimental parameters are $V_\textup{pp}=0.275$~V, $V_\textup{PL}=-1.42$~V. (b) Measurements of pumped current as a function of $V_\textup{Bout}$ (circles) and fits to theoretical models (solid lines) for $V_\textup{Bin}=-0.601$~V. Other experimental parameters as in panel (a). Insets: selected data from the main panel on exploded scales to closely compare the goodness of the fits.}
\label{fig:fits}
\end{figure}

\begin{figure*}[t]
\includegraphics[scale=0.85]{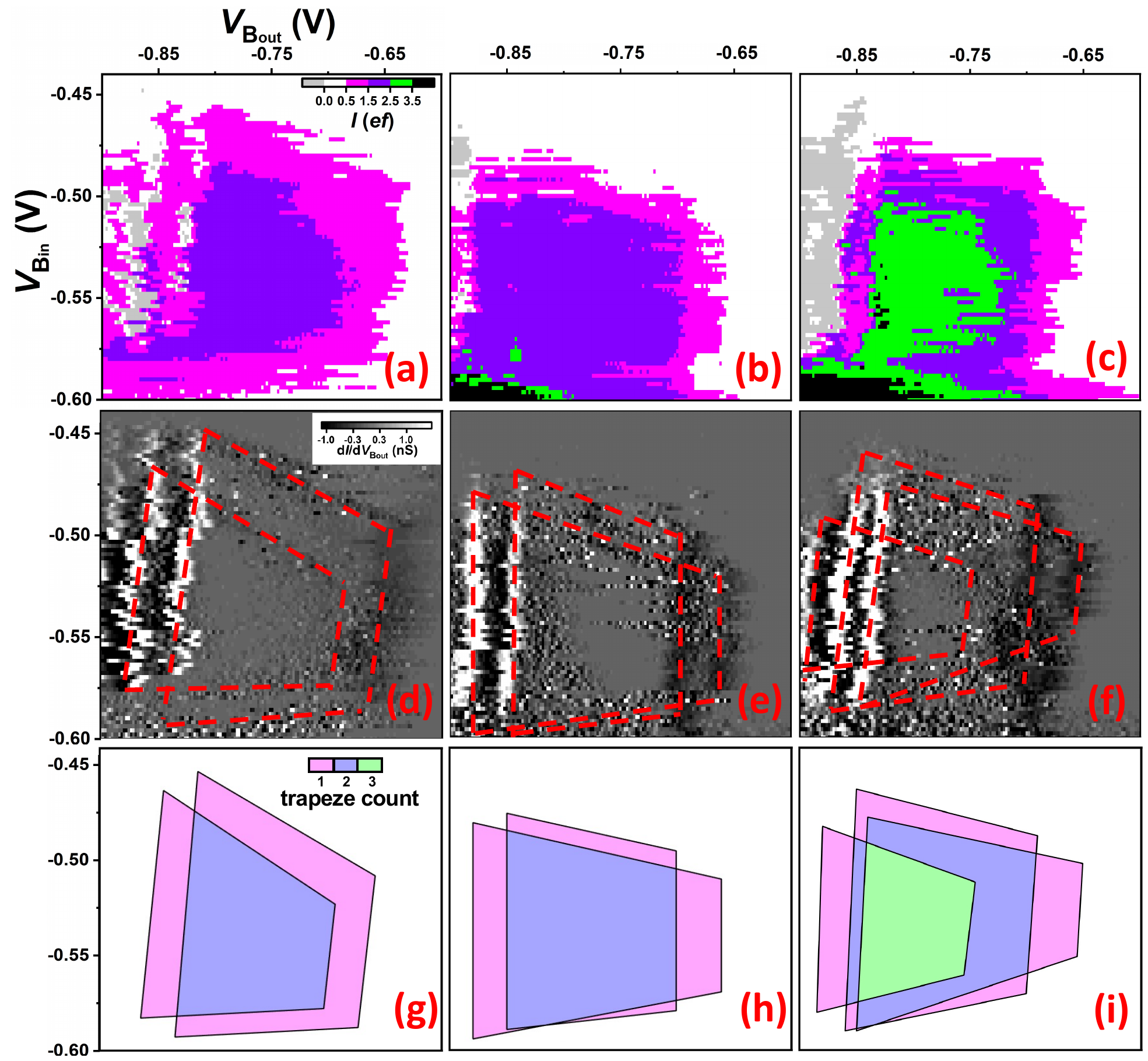}
\caption{Pumped current as a function of entrance and exit barrier gate voltages at $f=100$~MHz, $V_\textup{pp}=0.275$~V, (a) $V_\textup{PL}=-1.44$~V and $B=0$~T, (b) $V_\textup{PL}=-1.42$~V and $B=0$~T, (c) $V_\textup{PL}=-1.42$~V and $B=5$~T. Scale bar is the same for these three panels. (d) (e) (f) Derivative with respect to $V_\textup{Bout}$ of data in (a), (b) and (c), respectively. Scale bar is the same for these three panels. Red dashed lines are guides to the eye to highlight plateau boundaries. (g) (h) (i) Geometric trapezes drawn from the construction lines in (d), (e) and (f), respectively. A Boolean function is used to identify regions where trapeze areas overlap and assign different colors: magenta for no overlap, purple for overlap of two trapezes, green for overlap of three trapezes. Voltage ranges on x and y axis are the same for all nine panels. }
\label{fig:emul}
\end{figure*}

In order to tune the device into a single-QD operation regime, the transconductance is measured as a function of dc voltages applied to both barriers with the rf source turned off. As illustrated in Fig.~\ref{fig:dc}(d), for a fixed $V_\textup{PL}=-1.2$~V,  parallel coulomb blockade peaks appear diagonally across the studied parameter space, a clear indication of single QD formation~\cite{vanderwiel}. For less negative values of $V_\textup{PL}$, honeycomb-like stability diagrams are observed (not shown), suggesting a double QD regime instead. This informed the decision of carrying out ratchet experiments at $V_\textup{PL}<-1.2$~V. As highlighted by the dashed lines in panel (d), on occasions the coulomb peaks present abrupt discontinuities. This fact is an indication that random charge rearrangements are occurring in or in the vicinity of the QD, resulting in discrete jumps in the current level at a given operation point. It is likely that charge traps at the interface between different layers of the material stack may be responsible for this effect. Four (nominally identical) devices have been tested, and they have all showed roughly the same level of random fluctuations in d.c. tests.\\\indent
By applying a sinusoidal drive to the entrance gate with peak-to-peak voltage at the $50\Omega$ output of the source $V_\textup{pp}=0.275$~V, a current plateau at $I=ef$ emerges, as shown in Fig.~\ref{fig:freq}. For lower values of $V_\textup{pp}$ a current plateau is not observed. In fact, a strong capacitive coupling between the entrance gate and the QD in combination with a sufficiently large rf modulation provides captured holes with the energy shift needed to pass below the exit barrier and eventually be emitted at the end of a pump cycle~\cite{slava2010}, similarly to a previous report of silicon single-hole pumping~\cite{gento_hole}. In Fig.~\ref{fig:freq}(a), one can note the effect of the mentioned random fluctuations. The onset of the plateau region with respect to $V_\textup{Bout}$, the so-called capture line, undergoes discrete shifts at every voltage scan, as opposed to being linearly dependent on $V_\textup{Bin}$ as merely dictated by capacitive coupling considerations~\cite{Kaestner_2015}. By performing measurements at different frequencies we confirm that the pumped current scales as expected for a single hole transported per clock cycle, as reported in Fig.~\ref{fig:freq}(b). It is of note that in these measurements the current is not seen to increase towards $I=2ef$, as one would expect for increasingly negative exit barrier voltage, which would normally allow to trap and emit an additional hole. This may be due to a large charging energy in the QD, which prevents additional holes from being trapped, or insufficient amplitude of the rf drive resulting in incomplete emission~\cite{Kaestner_2015}.\\\indent
In order to test the latter hypothesis, we have increased the drive amplitude to $V_\textup{pp}=0.3~$V. However, this resulted in a severe worsening of the charge fluctuations and did not allow a conclusive test to be carried out. In order to modify the QD charging energy, a pump map is acquired at a more negative $V_\textup{PL}$, see Fig.~\ref{fig:fits}(a). This is expected to have the main effect of reducing the QD charging energy by enhancing its electrostatically-defined size. A secondary effect is the enhancement of both barriers transparency due to cross-coupling. The map shows that in these circumstances the current rises above the first plateau value for increasingly negative $V_\textup{Bout}$, albeit without fully reaching the second plateau.\\\indent
The shape of the current staircase between two adjacent
plateaux as a function of the exit barrier gate voltage can provide insights into the process
by which the QD is decoupled from the reservoir(s)~\cite{Kaestner_2015}, as indicated by the theoretical fits reported in Fig.~\ref{fig:fits}(b). Thus far, most of accurate semiconductor pumps~\cite{Giblin_2019} have operated in the decay cascade regime~\cite{slava2010}, where the final number of charged particles in the QD is determined by a one-way cascade of back-tunneling events. According to this model, the average number of holes captured per cycle can be written as 
\begin{equation}
\label{eq:decay}
    \langle n\rangle=\sum_{m}^{} \textup{exp[}-\textup{exp(}-\alpha V_\textup{Bout}+\Delta_m)]
\end{equation}
where $\alpha$ and $\Delta_m$ are fitting parameters. Alternatively, if the
reservoir in the vicinity of the entrance gate is heated by the large-amplitude ac
drive, charge capture may follow a thermal equilibrium regime. This operation mode has been previously observed in both electron and hole pumps in silicon~\cite{gento_hole,zhao}. In this regime, particles are exchanged between the QD and the leads only during the initial stage of the pumping cycle, and the average
number of captured holes can be written as~\cite{gento_thermal}
\begin{equation}
\label{eq:th-fit}
    \langle n\rangle=\sum_{m}^{} \frac{1}{1+\textup{exp(}A_m+BV_\textup{Bout})}
\end{equation}
where $A_m$ and $B$ are the fitting parameters for the $m$th current
plateau. In Fig.~\ref{fig:fits}(b), the normalized pumped current, $I/ef$, is used
in the numerical fit of $\langle n\rangle$ for both decay-cascade and
thermal models in the range $-0.95$~V~$\textless~V_\textup{Bout}\textless~-0.60$~V. A close inspection of the current staircase (see insets) shows that the thermal equilibrium model is a better fit than the decay cascade for the rising edge approaching $\langle n\rangle=1$. The fitting error for the thermal model, reduced-$\chi_\textup{th}^2=0.13$, is significantly
lower than for the decay cascade model, reduced-$\chi_\textup{decay}^2=0.85$, for
$V_\textup{Bin}=-0.601$~V. Similar results are obtained by fitting traces at other $V_\textup{Bin}$ values (not shown). This suggests that this hole pump operates in the thermal regime. However, it is clear that the rising edge to the second plateau is not well represented by either model. This is not unusual and it may be an indication that also other phenomena affect the pumping process, which may include a change in gate coupling between single- and multi-particle QD configurations~\cite{rossi2014}, as well as the presence of additional pumping entities such as traps or parasitic states besides the intended QD~\cite{gento-trap,rossi2018}.\\\indent
In order to investigate this aspect, pump maps showing multiple plateaux have been acquired under different experimental conditions, as shown in Fig.~\ref{fig:emul}. Maps in (a) and (b) are taken for different values of $V_\textup{PL}$ and both present current plateaux for $\langle n\rangle=1$ and $\langle n\rangle=2$ in a similar fashion to what is already shown in Fig.~\ref{fig:fits}(a). By contrast, the map in Fig.~\ref{fig:emul}(c) presents higher current values approaching $\langle n\rangle=3$ in addition to lower order plateaux. This measurement is taken with the device in a magnetic field, $B$, perpendicular to the hole layer plane and intensity $B=5$~T. The application of an out-of-plane magnetic field is expected to increase the QD confinement and improve the decoupling from the leads, which in GaAs pumps usually results in quantization enhancement~\cite{fletcher2012}. In a typical single-drive tunable-barrier pump the quantized plateau boundaries are set by insufficient loading or incomplete emission and form a checkers-type diagram with trapeze-shaped tiles at fixed values of $\langle n\rangle$~\cite{Kaestner_2015}. By contrast, the data presented here show that the regions of quantized current are nested one into another. This becomes clearer by looking at the pumped current derivatives shown in Fig.~\ref{fig:emul} (d), (e) and (f). The red construction lines highlight the boundaries of each plateau region and reveal significant overlaps between trapezes of different areas and orientation with respect to the gate space. This may suggest that multiple pumps are at work in this device, with each pump producing only a plateau at $\langle n\rangle=1$. As observed in Fig.~\ref{fig:freq}, this could be due to the fact that the QD is only able to capture one hole in the experimental conditions we have probed. By taking the Cartesian coordinates of the vertices of each trapeze, one can obtain their areas by simple geometric construction, as shown in Fig.~\ref{fig:emul} (g), (h), (i). A Boolean function is used to select regions of the 2D map space where the trapeze areas overlap and to assign different colors depending on how many overlaps are detected. Assuming that each trapeze sets the boundary for a different $\langle n\rangle=1$ plateau (magenta), the overlap of two trapezes would lead to a region of the map representative of $\langle n\rangle=2$ (purple), and similarly, three overlapping trapezes would lead to $\langle n\rangle=3$ (green). Comparing like-for-like panels in the bottom and top row of Fig.~\ref{fig:emul}, one may find enough similarities to indicate that multiple pumps operating in parallel could be the origin of the nested plateaux observed in the experiments.
\section{\label{sec:Disc}Discussion}
These experiments have been impacted by the frequent charge fluctuations in the device. The main limitation is that the effect of systematic changes in experimental parameters become intertwined with the effects of random charge rearrangements. Distinguishing  with confidence  between them is not always possible. Besides the instantaneous change of current levels that resulted in noisy pump maps, one has to account for a longer term instability caused by the accumulation in time of multiple random events. This becomes clear if one compares Fig.~\ref{fig:fits}(a) and Fig.~\ref{fig:emul}(b). These two maps are taken at the same experimental conditions (i.e. same drive frequency, amplitude, B-field and $V_\textup{PL}$ values) and yet the pumping regions appear radically different. This is likely due to the fact that the measurements had been taken few days apart.\\\indent
It is, therefore, out of an abundance of caution that we do not comment on the theoretical error rate of the pump, which one could have calculated based on the thermal fit value at the inflection point of the plateau~\cite{gento_hole,zhao}. Given that the plateau stability is affected by the fluctuators, it would be misleading to present such information for an isolated stable trace.\\\indent
In the absence of the random instabilities, one could have also tried to shed more light on the reasons for the capturing of just one hole in the QD and the origin of the multiple pumping mechanisms shown in Fig.~\ref{fig:emul}. For example, by systematically investigating pump maps as a function of magnetic and electrostatic confinements, one could have gathered relevant information on whether they originated from atomic-like trap states or parasitic QDs, depending on the effect on the plateau length and boundary.\\\indent
Finally, note that similar devices to those used in this work have resulted to be excellent hosts for spin-based quantum bits~\cite{nico2020}. In that case, stable charge configurations have been obtained by reducing gate voltage swings down to few mV. Unfortunately, this strategy does not lend itself effectively to pumping experiments where large sinusoidal drives are typically needed and extensive dc voltage scans have to be carried out to verify the robustness of the transfer protocol~\cite{Giblin_2017}.
\section{\label{sec:Conc}Conclusion}
In summary, we demonstrated a prototype of single-hole pump in a Ge-based QD. By fitting the current staircase to theoretical models, we conclude that the pump operates in a thermal regime. We observed unusual quantized plateaux boundaries in the 2D pump maps and attributed them to unintended parallel pump operation. More in depth analysis around the theoretical pump error rate, as well as the possible origin of the spurious pumping mechanisms was prevented by device instability in the form of random charge fluctuations.\\\indent
In the future, these pumps may become useful for metrological applications by contributing to high-accuracy current generation or universality tests. Furthermore, hole pumps in high-mobility Ge could be of interest for the nascent field of fermionic quantum optics~\cite{Bocquillon1054}, as well as for the realisation of single-photon sources based on charge transfer~\cite{imamoglu}. However, to fulfill these expectations, it will be imperative to improve their charge stability. Recent studies~\cite{Lodari_2021,nico2021} have shown  that quieter QDs can be realised when the Ge quantum well is positioned deeper in the heterostructure stack as a result of a thicker ($55$~nm) Si$_{0.2}$Ge$_{0.8}$ barrier layer. This also suggests that a possible origin of charge fluctuations resides in trap states at the interface between the Si capping layer and the barrier layer. We expect that the next generation of Ge pumps will directly benefit from this improvement in the fabrication process and will likely achieve much reduced levels of charge fluctuations.
\begin{acknowledgments}
We thank J. Fletcher for a critical reading of the manuscript, P. See for taking SEM images of the samples, and the other members of the Single Electron Team at the National Physical Laboratory for useful discussions. AR and MK acknowledge the support of the UK Government Department for Business, Energy and Industrial Strategy. AR also acknowledges support from a UKRI Future Leaders Fellowship (MR/T041110/1). GS and MV acknowledge support through an FOM Projectruimte of the Foundation for Fundamental Research on
Matter (FOM), associated with the Netherlands Organisation for Scientific Research (NWO).
\end{acknowledgments}
\bibliography{ref_main}

\providecommand{\noopsort}[1]{}\providecommand{\singleletter}[1]{#1}%
\begin{thebibliography}{33}%
\makeatletter
\providecommand \@ifxundefined [1]{%
 \@ifx{#1\undefined}
}%
\providecommand \@ifnum [1]{%
 \ifnum #1\expandafter \@firstoftwo
 \else \expandafter \@secondoftwo
 \fi
}%
\providecommand \@ifx [1]{%
 \ifx #1\expandafter \@firstoftwo
 \else \expandafter \@secondoftwo
 \fi
}%
\providecommand \natexlab [1]{#1}%
\providecommand \enquote  [1]{``#1''}%
\providecommand \bibnamefont  [1]{#1}%
\providecommand \bibfnamefont [1]{#1}%
\providecommand \citenamefont [1]{#1}%
\providecommand \href@noop [0]{\@secondoftwo}%
\providecommand \href [0]{\begingroup \@sanitize@url \@href}%
\providecommand \@href[1]{\@@startlink{#1}\@@href}%
\providecommand \@@href[1]{\endgroup#1\@@endlink}%
\providecommand \@sanitize@url [0]{\catcode `\\12\catcode `\$12\catcode
  `\&12\catcode `\#12\catcode `\^12\catcode `\_12\catcode `\%12\relax}%
\providecommand \@@startlink[1]{}%
\providecommand \@@endlink[0]{}%
\providecommand \url  [0]{\begingroup\@sanitize@url \@url }%
\providecommand \@url [1]{\endgroup\@href {#1}{\urlprefix }}%
\providecommand \urlprefix  [0]{URL }%
\providecommand \Eprint [0]{\href }%
\providecommand \doibase [0]{https://doi.org/}%
\providecommand \selectlanguage [0]{\@gobble}%
\providecommand \bibinfo  [0]{\@secondoftwo}%
\providecommand \bibfield  [0]{\@secondoftwo}%
\providecommand \translation [1]{[#1]}%
\providecommand \BibitemOpen [0]{}%
\providecommand \bibitemStop [0]{}%
\providecommand \bibitemNoStop [0]{.\EOS\space}%
\providecommand \EOS [0]{\spacefactor3000\relax}%
\providecommand \BibitemShut  [1]{\csname bibitem#1\endcsname}%
\let\auto@bib@innerbib\@empty
\bibitem [{\citenamefont {Pekola}\ \emph {et~al.}(2013)\citenamefont {Pekola},
  \citenamefont {Saira}, \citenamefont {Maisi}, \citenamefont {Kemppinen},
  \citenamefont {M\"ott\"onen}, \citenamefont {Pashkin},\ and\ \citenamefont
  {Averin}}]{pekola-rev}%
  \BibitemOpen
  \bibfield  {author} {\bibinfo {author} {\bibfnamefont {J.~P.}\ \bibnamefont
  {Pekola}}, \bibinfo {author} {\bibfnamefont {O.-P.}\ \bibnamefont {Saira}},
  \bibinfo {author} {\bibfnamefont {V.~F.}\ \bibnamefont {Maisi}}, \bibinfo
  {author} {\bibfnamefont {A.}~\bibnamefont {Kemppinen}}, \bibinfo {author}
  {\bibfnamefont {M.}~\bibnamefont {M\"ott\"onen}}, \bibinfo {author}
  {\bibfnamefont {Y.~A.}\ \bibnamefont {Pashkin}},\ and\ \bibinfo {author}
  {\bibfnamefont {D.~V.}\ \bibnamefont {Averin}},\ }\bibfield  {title}
  {\bibinfo {title} {Single-electron current sources: Toward a refined
  definition of the ampere},\ }\href
  {https://doi.org/10.1103/RevModPhys.85.1421} {\bibfield  {journal} {\bibinfo
  {journal} {Rev. Mod. Phys.}\ }\textbf {\bibinfo {volume} {85}},\ \bibinfo
  {pages} {1421} (\bibinfo {year} {2013})}\BibitemShut {NoStop}%
\bibitem [{\citenamefont {Scherer}\ and\ \citenamefont
  {Schumacher}(2019)}]{scherer-rev}%
  \BibitemOpen
  \bibfield  {author} {\bibinfo {author} {\bibfnamefont {H.}~\bibnamefont
  {Scherer}}\ and\ \bibinfo {author} {\bibfnamefont {H.~W.}\ \bibnamefont
  {Schumacher}},\ }\bibfield  {title} {\bibinfo {title} {Single-electron pumps
  and quantum current metrology in the {R}evised {SI}},\ }\href
  {https://doi.org/https://doi.org/10.1002/andp.201800371} {\bibfield
  {journal} {\bibinfo  {journal} {Annalen der Physik}\ }\textbf {\bibinfo
  {volume} {531}},\ \bibinfo {pages} {1800371} (\bibinfo {year}
  {2019})}\BibitemShut {NoStop}%
\bibitem [{\citenamefont {Zhao}\ \emph {et~al.}(2017)\citenamefont {Zhao},
  \citenamefont {Rossi}, \citenamefont {Giblin}, \citenamefont {Fletcher},
  \citenamefont {Hudson}, \citenamefont {M\"ott\"onen}, \citenamefont
  {Kataoka},\ and\ \citenamefont {Dzurak}}]{zhao}%
  \BibitemOpen
  \bibfield  {author} {\bibinfo {author} {\bibfnamefont {R.}~\bibnamefont
  {Zhao}}, \bibinfo {author} {\bibfnamefont {A.}~\bibnamefont {Rossi}},
  \bibinfo {author} {\bibfnamefont {S.~P.}\ \bibnamefont {Giblin}}, \bibinfo
  {author} {\bibfnamefont {J.~D.}\ \bibnamefont {Fletcher}}, \bibinfo {author}
  {\bibfnamefont {F.~E.}\ \bibnamefont {Hudson}}, \bibinfo {author}
  {\bibfnamefont {M.}~\bibnamefont {M\"ott\"onen}}, \bibinfo {author}
  {\bibfnamefont {M.}~\bibnamefont {Kataoka}},\ and\ \bibinfo {author}
  {\bibfnamefont {A.~S.}\ \bibnamefont {Dzurak}},\ }\bibfield  {title}
  {\bibinfo {title} {Thermal-error regime in high-accuracy gigahertz
  single-electron pumping},\ }\href
  {https://doi.org/10.1103/PhysRevApplied.8.044021} {\bibfield  {journal}
  {\bibinfo  {journal} {Phys. Rev. Applied}\ }\textbf {\bibinfo {volume} {8}},\
  \bibinfo {pages} {044021} (\bibinfo {year} {2017})}\BibitemShut {NoStop}%
\bibitem [{\citenamefont {Giblin}\ \emph {et~al.}(2019)\citenamefont {Giblin},
  \citenamefont {Fujiwara}, \citenamefont {Yamahata}, \citenamefont {Bae},
  \citenamefont {Kim}, \citenamefont {Rossi}, \citenamefont {M\"ott\"onen},\
  and\ \citenamefont {Kataoka}}]{Giblin_2019}%
  \BibitemOpen
  \bibfield  {author} {\bibinfo {author} {\bibfnamefont {S.~P.}\ \bibnamefont
  {Giblin}}, \bibinfo {author} {\bibfnamefont {A.}~\bibnamefont {Fujiwara}},
  \bibinfo {author} {\bibfnamefont {G.}~\bibnamefont {Yamahata}}, \bibinfo
  {author} {\bibfnamefont {M.-H.}\ \bibnamefont {Bae}}, \bibinfo {author}
  {\bibfnamefont {N.}~\bibnamefont {Kim}}, \bibinfo {author} {\bibfnamefont
  {A.}~\bibnamefont {Rossi}}, \bibinfo {author} {\bibfnamefont
  {M.}~\bibnamefont {M\"ott\"onen}},\ and\ \bibinfo {author} {\bibfnamefont
  {M.}~\bibnamefont {Kataoka}},\ }\bibfield  {title} {\bibinfo {title}
  {Evidence for universality of tunable-barrier electron pumps},\ }\href
  {https://doi.org/10.1088/1681-7575/ab29a5} {\bibfield  {journal} {\bibinfo
  {journal} {Metrologia}\ }\textbf {\bibinfo {volume} {56}},\ \bibinfo {pages}
  {044004} (\bibinfo {year} {2019})}\BibitemShut {NoStop}%
\bibitem [{\citenamefont {Yamahata}\ \emph {et~al.}(2016)\citenamefont
  {Yamahata}, \citenamefont {Giblin}, \citenamefont {Kataoka}, \citenamefont
  {Karasawa},\ and\ \citenamefont {Fujiwara}}]{gento-acc}%
  \BibitemOpen
  \bibfield  {author} {\bibinfo {author} {\bibfnamefont {G.}~\bibnamefont
  {Yamahata}}, \bibinfo {author} {\bibfnamefont {S.~P.}\ \bibnamefont
  {Giblin}}, \bibinfo {author} {\bibfnamefont {M.}~\bibnamefont {Kataoka}},
  \bibinfo {author} {\bibfnamefont {T.}~\bibnamefont {Karasawa}},\ and\
  \bibinfo {author} {\bibfnamefont {A.}~\bibnamefont {Fujiwara}},\ }\bibfield
  {title} {\bibinfo {title} {Gigahertz single-electron pumping in silicon with
  an accuracy better than 9.2 parts in 107},\ }\href
  {https://doi.org/10.1063/1.4953872} {\bibfield  {journal} {\bibinfo
  {journal} {Applied Physics Letters}\ }\textbf {\bibinfo {volume} {109}},\
  \bibinfo {pages} {013101} (\bibinfo {year} {2016})},\ \Eprint
  {https://arxiv.org/abs/https://doi.org/10.1063/1.4953872}
  {https://doi.org/10.1063/1.4953872} \BibitemShut {NoStop}%
\bibitem [{\citenamefont {Stein}\ \emph {et~al.}(2015)\citenamefont {Stein},
  \citenamefont {Drung}, \citenamefont {Fricke}, \citenamefont {Scherer},
  \citenamefont {Hohls}, \citenamefont {Leicht}, \citenamefont {G\"otz},
  \citenamefont {Krause}, \citenamefont {Behr}, \citenamefont {Pesel},
  \citenamefont {Pierz}, \citenamefont {Siegner}, \citenamefont {Ahlers},\ and\
  \citenamefont {Schumacher}}]{stein-2015}%
  \BibitemOpen
  \bibfield  {author} {\bibinfo {author} {\bibfnamefont {F.}~\bibnamefont
  {Stein}}, \bibinfo {author} {\bibfnamefont {D.}~\bibnamefont {Drung}},
  \bibinfo {author} {\bibfnamefont {L.}~\bibnamefont {Fricke}}, \bibinfo
  {author} {\bibfnamefont {H.}~\bibnamefont {Scherer}}, \bibinfo {author}
  {\bibfnamefont {F.}~\bibnamefont {Hohls}}, \bibinfo {author} {\bibfnamefont
  {C.}~\bibnamefont {Leicht}}, \bibinfo {author} {\bibfnamefont
  {M.}~\bibnamefont {G\"otz}}, \bibinfo {author} {\bibfnamefont
  {C.}~\bibnamefont {Krause}}, \bibinfo {author} {\bibfnamefont
  {R.}~\bibnamefont {Behr}}, \bibinfo {author} {\bibfnamefont {E.}~\bibnamefont
  {Pesel}}, \bibinfo {author} {\bibfnamefont {K.}~\bibnamefont {Pierz}},
  \bibinfo {author} {\bibfnamefont {U.}~\bibnamefont {Siegner}}, \bibinfo
  {author} {\bibfnamefont {F.~J.}\ \bibnamefont {Ahlers}},\ and\ \bibinfo
  {author} {\bibfnamefont {H.~W.}\ \bibnamefont {Schumacher}},\ }\bibfield
  {title} {\bibinfo {title} {Validation of a quantized-current source with 0.2
  ppm uncertainty},\ }\href {https://doi.org/10.1063/1.4930142} {\bibfield
  {journal} {\bibinfo  {journal} {Applied Physics Letters}\ }\textbf {\bibinfo
  {volume} {107}},\ \bibinfo {pages} {103501} (\bibinfo {year} {2015})},\
  \Eprint {https://arxiv.org/abs/https://doi.org/10.1063/1.4930142}
  {https://doi.org/10.1063/1.4930142} \BibitemShut {NoStop}%
\bibitem [{\citenamefont {Scherer}\ and\ \citenamefont
  {Camarota}(2012)}]{Scherer_2012}%
  \BibitemOpen
  \bibfield  {author} {\bibinfo {author} {\bibfnamefont {H.}~\bibnamefont
  {Scherer}}\ and\ \bibinfo {author} {\bibfnamefont {B.}~\bibnamefont
  {Camarota}},\ }\bibfield  {title} {\bibinfo {title} {Quantum metrology
  triangle experiments: a status review},\ }\href
  {https://doi.org/10.1088/0957-0233/23/12/124010} {\bibfield  {journal}
  {\bibinfo  {journal} {Measurement Science and Technology}\ }\textbf {\bibinfo
  {volume} {23}},\ \bibinfo {pages} {124010} (\bibinfo {year}
  {2012})}\BibitemShut {NoStop}%
\bibitem [{\citenamefont {Jeckelmann}\ \emph {et~al.}(1997)\citenamefont
  {Jeckelmann}, \citenamefont {Jeanneret},\ and\ \citenamefont
  {Inglis}}]{Si-hall}%
  \BibitemOpen
  \bibfield  {author} {\bibinfo {author} {\bibfnamefont {B.}~\bibnamefont
  {Jeckelmann}}, \bibinfo {author} {\bibfnamefont {B.}~\bibnamefont
  {Jeanneret}},\ and\ \bibinfo {author} {\bibfnamefont {D.}~\bibnamefont
  {Inglis}},\ }\bibfield  {title} {\bibinfo {title} {High-precision
  measurements of the quantized hall resistance:experimental conditions for
  universality},\ }\href {https://doi.org/10.1103/PhysRevB.55.13124} {\bibfield
   {journal} {\bibinfo  {journal} {Phys. Rev. B}\ }\textbf {\bibinfo {volume}
  {55}},\ \bibinfo {pages} {13124} (\bibinfo {year} {1997})}\BibitemShut
  {NoStop}%
\bibitem [{\citenamefont {Janssen}\ \emph {et~al.}(2012)\citenamefont
  {Janssen}, \citenamefont {Williams}, \citenamefont {Fletcher}, \citenamefont
  {Goebel}, \citenamefont {Tzalenchuk}, \citenamefont {Yakimova}, \citenamefont
  {Lara-Avila}, \citenamefont {Kubatkin},\ and\ \citenamefont
  {Fal{\textquotesingle}ko}}]{Janssen_2012}%
  \BibitemOpen
  \bibfield  {author} {\bibinfo {author} {\bibfnamefont {T.~J. B.~M.}\
  \bibnamefont {Janssen}}, \bibinfo {author} {\bibfnamefont {J.~M.}\
  \bibnamefont {Williams}}, \bibinfo {author} {\bibfnamefont {N.~E.}\
  \bibnamefont {Fletcher}}, \bibinfo {author} {\bibfnamefont {R.}~\bibnamefont
  {Goebel}}, \bibinfo {author} {\bibfnamefont {A.}~\bibnamefont {Tzalenchuk}},
  \bibinfo {author} {\bibfnamefont {R.}~\bibnamefont {Yakimova}}, \bibinfo
  {author} {\bibfnamefont {S.}~\bibnamefont {Lara-Avila}}, \bibinfo {author}
  {\bibfnamefont {S.}~\bibnamefont {Kubatkin}},\ and\ \bibinfo {author}
  {\bibfnamefont {V.~I.}\ \bibnamefont {Fal{\textquotesingle}ko}},\ }\bibfield
  {title} {\bibinfo {title} {Precision comparison of the quantum hall effect in
  graphene and gallium arsenide},\ }\href
  {https://doi.org/10.1088/0026-1394/49/3/294} {\bibfield  {journal} {\bibinfo
  {journal} {Metrologia}\ }\textbf {\bibinfo {volume} {49}},\ \bibinfo {pages}
  {294} (\bibinfo {year} {2012})}\BibitemShut {NoStop}%
\bibitem [{\citenamefont {Niskanen}\ \emph {et~al.}(2005)\citenamefont
  {Niskanen}, \citenamefont {Kivioja}, \citenamefont {Sepp\"a},\ and\
  \citenamefont {Pekola}}]{cooper1}%
  \BibitemOpen
  \bibfield  {author} {\bibinfo {author} {\bibfnamefont {A.~O.}\ \bibnamefont
  {Niskanen}}, \bibinfo {author} {\bibfnamefont {J.~M.}\ \bibnamefont
  {Kivioja}}, \bibinfo {author} {\bibfnamefont {H.}~\bibnamefont {Sepp\"a}},\
  and\ \bibinfo {author} {\bibfnamefont {J.~P.}\ \bibnamefont {Pekola}},\
  }\bibfield  {title} {\bibinfo {title} {Evidence of cooper-pair pumping with
  combined flux and voltage control},\ }\href
  {https://doi.org/10.1103/PhysRevB.71.012513} {\bibfield  {journal} {\bibinfo
  {journal} {Phys. Rev. B}\ }\textbf {\bibinfo {volume} {71}},\ \bibinfo
  {pages} {012513} (\bibinfo {year} {2005})}\BibitemShut {NoStop}%
\bibitem [{\citenamefont {Vartiainen}\ \emph {et~al.}(2007)\citenamefont
  {Vartiainen}, \citenamefont {M\"ott\"onen}, \citenamefont {Pekola},\ and\
  \citenamefont {Kemppinen}}]{cooper2}%
  \BibitemOpen
  \bibfield  {author} {\bibinfo {author} {\bibfnamefont {J.~J.}\ \bibnamefont
  {Vartiainen}}, \bibinfo {author} {\bibfnamefont {M.}~\bibnamefont
  {M\"ott\"onen}}, \bibinfo {author} {\bibfnamefont {J.~P.}\ \bibnamefont
  {Pekola}},\ and\ \bibinfo {author} {\bibfnamefont {A.}~\bibnamefont
  {Kemppinen}},\ }\bibfield  {title} {\bibinfo {title} {Nanoampere pumping of
  cooper pairs},\ }\href {https://doi.org/10.1063/1.2709967} {\bibfield
  {journal} {\bibinfo  {journal} {Applied Physics Letters}\ }\textbf {\bibinfo
  {volume} {90}},\ \bibinfo {pages} {082102} (\bibinfo {year} {2007})},\
  \Eprint {https://arxiv.org/abs/https://doi.org/10.1063/1.2709967}
  {https://doi.org/10.1063/1.2709967} \BibitemShut {NoStop}%
\bibitem [{\citenamefont {Pothier}\ \emph {et~al.}(1992)\citenamefont
  {Pothier}, \citenamefont {Lafarge}, \citenamefont {Urbina}, \citenamefont
  {Esteve},\ and\ \citenamefont {Devoret}}]{Pothier_1992}%
  \BibitemOpen
  \bibfield  {author} {\bibinfo {author} {\bibfnamefont {H.}~\bibnamefont
  {Pothier}}, \bibinfo {author} {\bibfnamefont {P.}~\bibnamefont {Lafarge}},
  \bibinfo {author} {\bibfnamefont {C.}~\bibnamefont {Urbina}}, \bibinfo
  {author} {\bibfnamefont {D.}~\bibnamefont {Esteve}},\ and\ \bibinfo {author}
  {\bibfnamefont {M.~H.}\ \bibnamefont {Devoret}},\ }\bibfield  {title}
  {\bibinfo {title} {Single-electron pump based on charging effects},\ }\href
  {https://doi.org/10.1209/0295-5075/17/3/011} {\bibfield  {journal} {\bibinfo
  {journal} {Europhysics Letters ({EPL})}\ }\textbf {\bibinfo {volume} {17}},\
  \bibinfo {pages} {249} (\bibinfo {year} {1992})}\BibitemShut {NoStop}%
\bibitem [{\citenamefont {Pekola}\ \emph {et~al.}(2008)\citenamefont {Pekola},
  \citenamefont {Vartiainen}, \citenamefont {M{\"o}tt{\"o}nen}, \citenamefont
  {Saira}, \citenamefont {Meschke},\ and\ \citenamefont {Averin}}]{hybrid}%
  \BibitemOpen
  \bibfield  {author} {\bibinfo {author} {\bibfnamefont {J.~P.}\ \bibnamefont
  {Pekola}}, \bibinfo {author} {\bibfnamefont {J.~J.}\ \bibnamefont
  {Vartiainen}}, \bibinfo {author} {\bibfnamefont {M.}~\bibnamefont
  {M{\"o}tt{\"o}nen}}, \bibinfo {author} {\bibfnamefont {O.-P.}\ \bibnamefont
  {Saira}}, \bibinfo {author} {\bibfnamefont {M.}~\bibnamefont {Meschke}},\
  and\ \bibinfo {author} {\bibfnamefont {D.~V.}\ \bibnamefont {Averin}},\
  }\bibfield  {title} {\bibinfo {title} {Hybrid single-electron transistor as a
  source of quantized electric current},\ }\href
  {https://doi.org/10.1038/nphys808} {\bibfield  {journal} {\bibinfo  {journal}
  {Nature Physics}\ }\textbf {\bibinfo {volume} {4}},\ \bibinfo {pages} {120}
  (\bibinfo {year} {2008})}\BibitemShut {NoStop}%
\bibitem [{\citenamefont {Tettamanzi}\ \emph {et~al.}(2014)\citenamefont
  {Tettamanzi}, \citenamefont {Wacquez},\ and\ \citenamefont
  {Rogge}}]{Tettamanzi_2014}%
  \BibitemOpen
  \bibfield  {author} {\bibinfo {author} {\bibfnamefont {G.~C.}\ \bibnamefont
  {Tettamanzi}}, \bibinfo {author} {\bibfnamefont {R.}~\bibnamefont
  {Wacquez}},\ and\ \bibinfo {author} {\bibfnamefont {S.}~\bibnamefont
  {Rogge}},\ }\bibfield  {title} {\bibinfo {title} {Charge pumping through a
  single donor atom},\ }\href {https://doi.org/10.1088/1367-2630/16/6/063036}
  {\bibfield  {journal} {\bibinfo  {journal} {New Journal of Physics}\ }\textbf
  {\bibinfo {volume} {16}},\ \bibinfo {pages} {063036} (\bibinfo {year}
  {2014})}\BibitemShut {NoStop}%
\bibitem [{\citenamefont {Roche}\ \emph {et~al.}(2013)\citenamefont {Roche},
  \citenamefont {Riwar}, \citenamefont {Voisin}, \citenamefont
  {Dupont-Ferrier}, \citenamefont {Wacquez}, \citenamefont {Vinet},
  \citenamefont {Sanquer}, \citenamefont {Splettstoesser},\ and\ \citenamefont
  {Jehl}}]{roche}%
  \BibitemOpen
  \bibfield  {author} {\bibinfo {author} {\bibfnamefont {B.}~\bibnamefont
  {Roche}}, \bibinfo {author} {\bibfnamefont {R.~P.}\ \bibnamefont {Riwar}},
  \bibinfo {author} {\bibfnamefont {B.}~\bibnamefont {Voisin}}, \bibinfo
  {author} {\bibfnamefont {E.}~\bibnamefont {Dupont-Ferrier}}, \bibinfo
  {author} {\bibfnamefont {R.}~\bibnamefont {Wacquez}}, \bibinfo {author}
  {\bibfnamefont {M.}~\bibnamefont {Vinet}}, \bibinfo {author} {\bibfnamefont
  {M.}~\bibnamefont {Sanquer}}, \bibinfo {author} {\bibfnamefont
  {J.}~\bibnamefont {Splettstoesser}},\ and\ \bibinfo {author} {\bibfnamefont
  {X.}~\bibnamefont {Jehl}},\ }\bibfield  {title} {\bibinfo {title} {A two-atom
  electron pump},\ }\href {https://doi.org/10.1038/ncomms2544} {\bibfield
  {journal} {\bibinfo  {journal} {Nature Communications}\ }\textbf {\bibinfo
  {volume} {4}},\ \bibinfo {pages} {1581} (\bibinfo {year} {2013})}\BibitemShut
  {NoStop}%
\bibitem [{\citenamefont {Connolly}\ \emph {et~al.}(2013)\citenamefont
  {Connolly}, \citenamefont {Chiu}, \citenamefont {Giblin}, \citenamefont
  {Kataoka}, \citenamefont {Fletcher}, \citenamefont {Chua}, \citenamefont
  {Griffiths}, \citenamefont {Jones}, \citenamefont {Fal'ko}, \citenamefont
  {Smith},\ and\ \citenamefont {Janssen}}]{graphene}%
  \BibitemOpen
  \bibfield  {author} {\bibinfo {author} {\bibfnamefont {M.~R.}\ \bibnamefont
  {Connolly}}, \bibinfo {author} {\bibfnamefont {K.~L.}\ \bibnamefont {Chiu}},
  \bibinfo {author} {\bibfnamefont {S.~P.}\ \bibnamefont {Giblin}}, \bibinfo
  {author} {\bibfnamefont {M.}~\bibnamefont {Kataoka}}, \bibinfo {author}
  {\bibfnamefont {J.~D.}\ \bibnamefont {Fletcher}}, \bibinfo {author}
  {\bibfnamefont {C.}~\bibnamefont {Chua}}, \bibinfo {author} {\bibfnamefont
  {J.~P.}\ \bibnamefont {Griffiths}}, \bibinfo {author} {\bibfnamefont
  {G.~A.~C.}\ \bibnamefont {Jones}}, \bibinfo {author} {\bibfnamefont {V.~I.}\
  \bibnamefont {Fal'ko}}, \bibinfo {author} {\bibfnamefont {C.~G.}\
  \bibnamefont {Smith}},\ and\ \bibinfo {author} {\bibfnamefont {T.~J. B.~M.}\
  \bibnamefont {Janssen}},\ }\bibfield  {title} {\bibinfo {title} {Gigahertz
  quantized charge pumping in graphene quantum dots},\ }\href
  {https://doi.org/10.1038/nnano.2013.73} {\bibfield  {journal} {\bibinfo
  {journal} {Nature Nanotechnology}\ }\textbf {\bibinfo {volume} {8}},\
  \bibinfo {pages} {417} (\bibinfo {year} {2013})}\BibitemShut {NoStop}%
\bibitem [{\citenamefont {Sammak}\ \emph {et~al.}(2019)\citenamefont {Sammak},
  \citenamefont {Sabbagh}, \citenamefont {Hendrickx}, \citenamefont {Lodari},
  \citenamefont {Paquelet~Wuetz}, \citenamefont {Tosato}, \citenamefont {Yeoh},
  \citenamefont {Bollani}, \citenamefont {Virgilio}, \citenamefont {Schubert},
  \citenamefont {Zaumseil}, \citenamefont {Capellini}, \citenamefont
  {Veldhorst},\ and\ \citenamefont {Scappucci}}]{amir2019}%
  \BibitemOpen
  \bibfield  {author} {\bibinfo {author} {\bibfnamefont {A.}~\bibnamefont
  {Sammak}}, \bibinfo {author} {\bibfnamefont {D.}~\bibnamefont {Sabbagh}},
  \bibinfo {author} {\bibfnamefont {N.~W.}\ \bibnamefont {Hendrickx}}, \bibinfo
  {author} {\bibfnamefont {M.}~\bibnamefont {Lodari}}, \bibinfo {author}
  {\bibfnamefont {B.}~\bibnamefont {Paquelet~Wuetz}}, \bibinfo {author}
  {\bibfnamefont {A.}~\bibnamefont {Tosato}}, \bibinfo {author} {\bibfnamefont
  {L.}~\bibnamefont {Yeoh}}, \bibinfo {author} {\bibfnamefont {M.}~\bibnamefont
  {Bollani}}, \bibinfo {author} {\bibfnamefont {M.}~\bibnamefont {Virgilio}},
  \bibinfo {author} {\bibfnamefont {M.~A.}\ \bibnamefont {Schubert}}, \bibinfo
  {author} {\bibfnamefont {P.}~\bibnamefont {Zaumseil}}, \bibinfo {author}
  {\bibfnamefont {G.}~\bibnamefont {Capellini}}, \bibinfo {author}
  {\bibfnamefont {M.}~\bibnamefont {Veldhorst}},\ and\ \bibinfo {author}
  {\bibfnamefont {G.}~\bibnamefont {Scappucci}},\ }\bibfield  {title} {\bibinfo
  {title} {Shallow and undoped germanium quantum wells: A playground for spin
  and hybrid quantum technology},\ }\href
  {https://doi.org/https://doi.org/10.1002/adfm.201807613} {\bibfield
  {journal} {\bibinfo  {journal} {Advanced Functional Materials}\ }\textbf
  {\bibinfo {volume} {29}},\ \bibinfo {pages} {1807613} (\bibinfo {year}
  {2019})}\BibitemShut {NoStop}%
\bibitem [{\citenamefont {Lodari}\ \emph {et~al.}(2019)\citenamefont {Lodari},
  \citenamefont {Tosato}, \citenamefont {Sabbagh}, \citenamefont {Schubert},
  \citenamefont {Capellini}, \citenamefont {Sammak}, \citenamefont
  {Veldhorst},\ and\ \citenamefont {Scappucci}}]{lodari2019}%
  \BibitemOpen
  \bibfield  {author} {\bibinfo {author} {\bibfnamefont {M.}~\bibnamefont
  {Lodari}}, \bibinfo {author} {\bibfnamefont {A.}~\bibnamefont {Tosato}},
  \bibinfo {author} {\bibfnamefont {D.}~\bibnamefont {Sabbagh}}, \bibinfo
  {author} {\bibfnamefont {M.~A.}\ \bibnamefont {Schubert}}, \bibinfo {author}
  {\bibfnamefont {G.}~\bibnamefont {Capellini}}, \bibinfo {author}
  {\bibfnamefont {A.}~\bibnamefont {Sammak}}, \bibinfo {author} {\bibfnamefont
  {M.}~\bibnamefont {Veldhorst}},\ and\ \bibinfo {author} {\bibfnamefont
  {G.}~\bibnamefont {Scappucci}},\ }\bibfield  {title} {\bibinfo {title} {Light
  effective hole mass in undoped ge/sige quantum wells},\ }\href
  {https://doi.org/10.1103/PhysRevB.100.041304} {\bibfield  {journal} {\bibinfo
   {journal} {Phys. Rev. B}\ }\textbf {\bibinfo {volume} {100}},\ \bibinfo
  {pages} {041304} (\bibinfo {year} {2019})}\BibitemShut {NoStop}%
\bibitem [{\citenamefont {Hendrickx}\ \emph {et~al.}(2020)\citenamefont
  {Hendrickx}, \citenamefont {Lawrie}, \citenamefont {Petit}, \citenamefont
  {Sammak}, \citenamefont {Scappucci},\ and\ \citenamefont
  {Veldhorst}}]{nico2020}%
  \BibitemOpen
  \bibfield  {author} {\bibinfo {author} {\bibfnamefont {N.~W.}\ \bibnamefont
  {Hendrickx}}, \bibinfo {author} {\bibfnamefont {W.~I.~L.}\ \bibnamefont
  {Lawrie}}, \bibinfo {author} {\bibfnamefont {L.}~\bibnamefont {Petit}},
  \bibinfo {author} {\bibfnamefont {A.}~\bibnamefont {Sammak}}, \bibinfo
  {author} {\bibfnamefont {G.}~\bibnamefont {Scappucci}},\ and\ \bibinfo
  {author} {\bibfnamefont {M.}~\bibnamefont {Veldhorst}},\ }\bibfield  {title}
  {\bibinfo {title} {A single-hole spin qubit},\ }\href
  {https://doi.org/10.1038/s41467-020-17211-7} {\bibfield  {journal} {\bibinfo
  {journal} {Nature Communications}\ }\textbf {\bibinfo {volume} {11}},\
  \bibinfo {pages} {3478} (\bibinfo {year} {2020})}\BibitemShut {NoStop}%
\bibitem [{\citenamefont {Kaestner}\ and\ \citenamefont
  {Kashcheyevs}(2015)}]{Kaestner_2015}%
  \BibitemOpen
  \bibfield  {author} {\bibinfo {author} {\bibfnamefont {B.}~\bibnamefont
  {Kaestner}}\ and\ \bibinfo {author} {\bibfnamefont {V.}~\bibnamefont
  {Kashcheyevs}},\ }\bibfield  {title} {\bibinfo {title} {Non-adiabatic
  quantized charge pumping with tunable-barrier quantum dots: a review of
  current progress},\ }\href {https://doi.org/10.1088/0034-4885/78/10/103901}
  {\bibfield  {journal} {\bibinfo  {journal} {Reports on Progress in Physics}\
  }\textbf {\bibinfo {volume} {78}},\ \bibinfo {pages} {103901} (\bibinfo
  {year} {2015})}\BibitemShut {NoStop}%
\bibitem [{\citenamefont {van~der Wiel}\ \emph {et~al.}(2002)\citenamefont
  {van~der Wiel}, \citenamefont {De~Franceschi}, \citenamefont {Elzerman},
  \citenamefont {Fujisawa}, \citenamefont {Tarucha},\ and\ \citenamefont
  {Kouwenhoven}}]{vanderwiel}%
  \BibitemOpen
  \bibfield  {author} {\bibinfo {author} {\bibfnamefont {W.~G.}\ \bibnamefont
  {van~der Wiel}}, \bibinfo {author} {\bibfnamefont {S.}~\bibnamefont
  {De~Franceschi}}, \bibinfo {author} {\bibfnamefont {J.~M.}\ \bibnamefont
  {Elzerman}}, \bibinfo {author} {\bibfnamefont {T.}~\bibnamefont {Fujisawa}},
  \bibinfo {author} {\bibfnamefont {S.}~\bibnamefont {Tarucha}},\ and\ \bibinfo
  {author} {\bibfnamefont {L.~P.}\ \bibnamefont {Kouwenhoven}},\ }\bibfield
  {title} {\bibinfo {title} {Electron transport through double quantum dots},\
  }\href {https://doi.org/10.1103/RevModPhys.75.1} {\bibfield  {journal}
  {\bibinfo  {journal} {Rev. Mod. Phys.}\ }\textbf {\bibinfo {volume} {75}},\
  \bibinfo {pages} {1} (\bibinfo {year} {2002})}\BibitemShut {NoStop}%
\bibitem [{\citenamefont {Kashcheyevs}\ and\ \citenamefont
  {Kaestner}(2010)}]{slava2010}%
  \BibitemOpen
  \bibfield  {author} {\bibinfo {author} {\bibfnamefont {V.}~\bibnamefont
  {Kashcheyevs}}\ and\ \bibinfo {author} {\bibfnamefont {B.}~\bibnamefont
  {Kaestner}},\ }\bibfield  {title} {\bibinfo {title} {Universal decay cascade
  model for dynamic quantum dot initialization},\ }\href
  {https://doi.org/10.1103/PhysRevLett.104.186805} {\bibfield  {journal}
  {\bibinfo  {journal} {Phys. Rev. Lett.}\ }\textbf {\bibinfo {volume} {104}},\
  \bibinfo {pages} {186805} (\bibinfo {year} {2010})}\BibitemShut {NoStop}%
\bibitem [{\citenamefont {Yamahata}\ \emph {et~al.}(2015)\citenamefont
  {Yamahata}, \citenamefont {Karasawa},\ and\ \citenamefont
  {Fujiwara}}]{gento_hole}%
  \BibitemOpen
  \bibfield  {author} {\bibinfo {author} {\bibfnamefont {G.}~\bibnamefont
  {Yamahata}}, \bibinfo {author} {\bibfnamefont {T.}~\bibnamefont {Karasawa}},\
  and\ \bibinfo {author} {\bibfnamefont {A.}~\bibnamefont {Fujiwara}},\
  }\bibfield  {title} {\bibinfo {title} {Gigahertz single-hole transfer in si
  tunable-barrier pumps},\ }\href {https://doi.org/10.1063/1.4905934}
  {\bibfield  {journal} {\bibinfo  {journal} {Applied Physics Letters}\
  }\textbf {\bibinfo {volume} {106}},\ \bibinfo {pages} {023112} (\bibinfo
  {year} {2015})},\ \Eprint
  {https://arxiv.org/abs/https://doi.org/10.1063/1.4905934}
  {https://doi.org/10.1063/1.4905934} \BibitemShut {NoStop}%
\bibitem [{\citenamefont {Yamahata}\ \emph
  {et~al.}(2014{\natexlab{a}})\citenamefont {Yamahata}, \citenamefont
  {Nishiguchi},\ and\ \citenamefont {Fujiwara}}]{gento_thermal}%
  \BibitemOpen
  \bibfield  {author} {\bibinfo {author} {\bibfnamefont {G.}~\bibnamefont
  {Yamahata}}, \bibinfo {author} {\bibfnamefont {K.}~\bibnamefont
  {Nishiguchi}},\ and\ \bibinfo {author} {\bibfnamefont {A.}~\bibnamefont
  {Fujiwara}},\ }\bibfield  {title} {\bibinfo {title} {Accuracy evaluation and
  mechanism crossover of single-electron transfer in si tunable-barrier
  turnstiles},\ }\href {https://doi.org/10.1103/PhysRevB.89.165302} {\bibfield
  {journal} {\bibinfo  {journal} {Phys. Rev. B}\ }\textbf {\bibinfo {volume}
  {89}},\ \bibinfo {pages} {165302} (\bibinfo {year}
  {2014}{\natexlab{a}})}\BibitemShut {NoStop}%
\bibitem [{\citenamefont {Rossi}\ \emph {et~al.}(2014)\citenamefont {Rossi},
  \citenamefont {Tanttu}, \citenamefont {Tan}, \citenamefont {Iisakka},
  \citenamefont {Zhao}, \citenamefont {Chan}, \citenamefont {Tettamanzi},
  \citenamefont {Rogge}, \citenamefont {Dzurak},\ and\ \citenamefont
  {M{\"o}tt{\"o}nen}}]{rossi2014}%
  \BibitemOpen
  \bibfield  {author} {\bibinfo {author} {\bibfnamefont {A.}~\bibnamefont
  {Rossi}}, \bibinfo {author} {\bibfnamefont {T.}~\bibnamefont {Tanttu}},
  \bibinfo {author} {\bibfnamefont {K.~Y.}\ \bibnamefont {Tan}}, \bibinfo
  {author} {\bibfnamefont {I.}~\bibnamefont {Iisakka}}, \bibinfo {author}
  {\bibfnamefont {R.}~\bibnamefont {Zhao}}, \bibinfo {author} {\bibfnamefont
  {K.~W.}\ \bibnamefont {Chan}}, \bibinfo {author} {\bibfnamefont {G.~C.}\
  \bibnamefont {Tettamanzi}}, \bibinfo {author} {\bibfnamefont
  {S.}~\bibnamefont {Rogge}}, \bibinfo {author} {\bibfnamefont {A.~S.}\
  \bibnamefont {Dzurak}},\ and\ \bibinfo {author} {\bibfnamefont
  {M.}~\bibnamefont {M{\"o}tt{\"o}nen}},\ }\bibfield  {title} {\bibinfo {title}
  {An accurate single-electron pump based on a highly tunable silicon quantum
  dot},\ }\href {https://doi.org/10.1021/nl500927q} {\bibfield  {journal}
  {\bibinfo  {journal} {Nano Letters}\ }\textbf {\bibinfo {volume} {14}},\
  \bibinfo {pages} {3405} (\bibinfo {year} {2014})}\BibitemShut {NoStop}%
\bibitem [{\citenamefont {Yamahata}\ \emph
  {et~al.}(2014{\natexlab{b}})\citenamefont {Yamahata}, \citenamefont
  {Nishiguchi},\ and\ \citenamefont {Fujiwara}}]{gento-trap}%
  \BibitemOpen
  \bibfield  {author} {\bibinfo {author} {\bibfnamefont {G.}~\bibnamefont
  {Yamahata}}, \bibinfo {author} {\bibfnamefont {K.}~\bibnamefont
  {Nishiguchi}},\ and\ \bibinfo {author} {\bibfnamefont {A.}~\bibnamefont
  {Fujiwara}},\ }\bibfield  {title} {\bibinfo {title} {Gigahertz single-trap
  electron pumps in silicon},\ }\href {https://doi.org/10.1038/ncomms6038}
  {\bibfield  {journal} {\bibinfo  {journal} {Nature Communications}\ }\textbf
  {\bibinfo {volume} {5}},\ \bibinfo {pages} {5038} (\bibinfo {year}
  {2014}{\natexlab{b}})}\BibitemShut {NoStop}%
\bibitem [{\citenamefont {Rossi}\ \emph {et~al.}(2018)\citenamefont {Rossi},
  \citenamefont {Klochan}, \citenamefont {Timoshenko}, \citenamefont {Hudson},
  \citenamefont {M{\"o}tt{\"o}nen}, \citenamefont {Rogge}, \citenamefont
  {Dzurak}, \citenamefont {Kashcheyevs},\ and\ \citenamefont
  {Tettamanzi}}]{rossi2018}%
  \BibitemOpen
  \bibfield  {author} {\bibinfo {author} {\bibfnamefont {A.}~\bibnamefont
  {Rossi}}, \bibinfo {author} {\bibfnamefont {J.}~\bibnamefont {Klochan}},
  \bibinfo {author} {\bibfnamefont {J.}~\bibnamefont {Timoshenko}}, \bibinfo
  {author} {\bibfnamefont {F.~E.}\ \bibnamefont {Hudson}}, \bibinfo {author}
  {\bibfnamefont {M.}~\bibnamefont {M{\"o}tt{\"o}nen}}, \bibinfo {author}
  {\bibfnamefont {S.}~\bibnamefont {Rogge}}, \bibinfo {author} {\bibfnamefont
  {A.~S.}\ \bibnamefont {Dzurak}}, \bibinfo {author} {\bibfnamefont
  {V.}~\bibnamefont {Kashcheyevs}},\ and\ \bibinfo {author} {\bibfnamefont
  {G.~C.}\ \bibnamefont {Tettamanzi}},\ }\bibfield  {title} {\bibinfo {title}
  {Gigahertz single-electron pumping mediated by parasitic states},\ }\href
  {https://doi.org/10.1021/acs.nanolett.8b00874} {\bibfield  {journal}
  {\bibinfo  {journal} {Nano Letters}\ }\textbf {\bibinfo {volume} {18}},\
  \bibinfo {pages} {4141} (\bibinfo {year} {2018})}\BibitemShut {NoStop}%
\bibitem [{\citenamefont {Fletcher}\ \emph {et~al.}(2012)\citenamefont
  {Fletcher}, \citenamefont {Kataoka}, \citenamefont {Giblin}, \citenamefont
  {Park}, \citenamefont {Sim}, \citenamefont {See}, \citenamefont {Ritchie},
  \citenamefont {Griffiths}, \citenamefont {Jones}, \citenamefont {Beere},\
  and\ \citenamefont {Janssen}}]{fletcher2012}%
  \BibitemOpen
  \bibfield  {author} {\bibinfo {author} {\bibfnamefont {J.~D.}\ \bibnamefont
  {Fletcher}}, \bibinfo {author} {\bibfnamefont {M.}~\bibnamefont {Kataoka}},
  \bibinfo {author} {\bibfnamefont {S.~P.}\ \bibnamefont {Giblin}}, \bibinfo
  {author} {\bibfnamefont {S.}~\bibnamefont {Park}}, \bibinfo {author}
  {\bibfnamefont {H.-S.}\ \bibnamefont {Sim}}, \bibinfo {author} {\bibfnamefont
  {P.}~\bibnamefont {See}}, \bibinfo {author} {\bibfnamefont {D.~A.}\
  \bibnamefont {Ritchie}}, \bibinfo {author} {\bibfnamefont {J.~P.}\
  \bibnamefont {Griffiths}}, \bibinfo {author} {\bibfnamefont {G.~A.~C.}\
  \bibnamefont {Jones}}, \bibinfo {author} {\bibfnamefont {H.~E.}\ \bibnamefont
  {Beere}},\ and\ \bibinfo {author} {\bibfnamefont {T.~J. B.~M.}\ \bibnamefont
  {Janssen}},\ }\bibfield  {title} {\bibinfo {title} {Stabilization of
  single-electron pumps by high magnetic fields},\ }\href
  {https://doi.org/10.1103/PhysRevB.86.155311} {\bibfield  {journal} {\bibinfo
  {journal} {Phys. Rev. B}\ }\textbf {\bibinfo {volume} {86}},\ \bibinfo
  {pages} {155311} (\bibinfo {year} {2012})}\BibitemShut {NoStop}%
\bibitem [{\citenamefont {Giblin}\ \emph {et~al.}(2017)\citenamefont {Giblin},
  \citenamefont {Bae}, \citenamefont {Kim}, \citenamefont {Ahn},\ and\
  \citenamefont {Kataoka}}]{Giblin_2017}%
  \BibitemOpen
  \bibfield  {author} {\bibinfo {author} {\bibfnamefont {S.~P.}\ \bibnamefont
  {Giblin}}, \bibinfo {author} {\bibfnamefont {M.-H.}\ \bibnamefont {Bae}},
  \bibinfo {author} {\bibfnamefont {N.}~\bibnamefont {Kim}}, \bibinfo {author}
  {\bibfnamefont {Y.-H.}\ \bibnamefont {Ahn}},\ and\ \bibinfo {author}
  {\bibfnamefont {M.}~\bibnamefont {Kataoka}},\ }\bibfield  {title} {\bibinfo
  {title} {Robust operation of a {GaAs} tunable barrier electron pump},\ }\href
  {https://doi.org/10.1088/1681-7575/aa634c} {\bibfield  {journal} {\bibinfo
  {journal} {Metrologia}\ }\textbf {\bibinfo {volume} {54}},\ \bibinfo {pages}
  {299} (\bibinfo {year} {2017})}\BibitemShut {NoStop}%
\bibitem [{\citenamefont {Bocquillon}\ \emph {et~al.}(2013)\citenamefont
  {Bocquillon}, \citenamefont {Freulon}, \citenamefont {Berroir}, \citenamefont
  {Degiovanni}, \citenamefont {Pla{\c c}ais}, \citenamefont {Cavanna},
  \citenamefont {Jin},\ and\ \citenamefont {F{\`e}ve}}]{Bocquillon1054}%
  \BibitemOpen
  \bibfield  {author} {\bibinfo {author} {\bibfnamefont {E.}~\bibnamefont
  {Bocquillon}}, \bibinfo {author} {\bibfnamefont {V.}~\bibnamefont {Freulon}},
  \bibinfo {author} {\bibfnamefont {J.-M.}\ \bibnamefont {Berroir}}, \bibinfo
  {author} {\bibfnamefont {P.}~\bibnamefont {Degiovanni}}, \bibinfo {author}
  {\bibfnamefont {B.}~\bibnamefont {Pla{\c c}ais}}, \bibinfo {author}
  {\bibfnamefont {A.}~\bibnamefont {Cavanna}}, \bibinfo {author} {\bibfnamefont
  {Y.}~\bibnamefont {Jin}},\ and\ \bibinfo {author} {\bibfnamefont
  {G.}~\bibnamefont {F{\`e}ve}},\ }\bibfield  {title} {\bibinfo {title}
  {Coherence and indistinguishability of single electrons emitted by
  independent sources},\ }\href {https://doi.org/10.1126/science.1232572}
  {\bibfield  {journal} {\bibinfo  {journal} {Science}\ }\textbf {\bibinfo
  {volume} {339}},\ \bibinfo {pages} {1054} (\bibinfo {year}
  {2013})}\BibitemShut {NoStop}%
\bibitem [{\citenamefont {Imamoglu}\ and\ \citenamefont
  {Yamamoto}(1994)}]{imamoglu}%
  \BibitemOpen
  \bibfield  {author} {\bibinfo {author} {\bibfnamefont {A.}~\bibnamefont
  {Imamoglu}}\ and\ \bibinfo {author} {\bibfnamefont {Y.}~\bibnamefont
  {Yamamoto}},\ }\bibfield  {title} {\bibinfo {title} {Turnstile device for
  heralded single photons: {C}oulomb blockade of electron and hole tunneling in
  quantum confined p-i-n heterojunctions},\ }\href
  {https://doi.org/10.1103/PhysRevLett.72.210} {\bibfield  {journal} {\bibinfo
  {journal} {Phys. Rev. Lett.}\ }\textbf {\bibinfo {volume} {72}},\ \bibinfo
  {pages} {210} (\bibinfo {year} {1994})}\BibitemShut {NoStop}%
\bibitem [{\citenamefont {Lodari}\ \emph {et~al.}(2021)\citenamefont {Lodari},
  \citenamefont {Hendrickx}, \citenamefont {Lawrie}, \citenamefont {Hsiao},
  \citenamefont {Vandersypen}, \citenamefont {Sammak}, \citenamefont
  {Veldhorst},\ and\ \citenamefont {Scappucci}}]{Lodari_2021}%
  \BibitemOpen
  \bibfield  {author} {\bibinfo {author} {\bibfnamefont {M.}~\bibnamefont
  {Lodari}}, \bibinfo {author} {\bibfnamefont {N.~W.}\ \bibnamefont
  {Hendrickx}}, \bibinfo {author} {\bibfnamefont {W.~I.~L.}\ \bibnamefont
  {Lawrie}}, \bibinfo {author} {\bibfnamefont {T.-K.}\ \bibnamefont {Hsiao}},
  \bibinfo {author} {\bibfnamefont {L.~M.~K.}\ \bibnamefont {Vandersypen}},
  \bibinfo {author} {\bibfnamefont {A.}~\bibnamefont {Sammak}}, \bibinfo
  {author} {\bibfnamefont {M.}~\bibnamefont {Veldhorst}},\ and\ \bibinfo
  {author} {\bibfnamefont {G.}~\bibnamefont {Scappucci}},\ }\bibfield  {title}
  {\bibinfo {title} {Low percolation density and charge noise with holes in
  germanium},\ }\href {https://doi.org/10.1088/2633-4356/abcd82} {\bibfield
  {journal} {\bibinfo  {journal} {Materials for Quantum Technology}\ }\textbf
  {\bibinfo {volume} {1}},\ \bibinfo {pages} {011002} (\bibinfo {year}
  {2021})}\BibitemShut {NoStop}%
\bibitem [{\citenamefont {Hendrickx}\ \emph {et~al.}(2021)\citenamefont
  {Hendrickx}, \citenamefont {Lawrie}, \citenamefont {Russ}, \citenamefont {van
  Riggelen}, \citenamefont {de~Snoo}, \citenamefont {Schouten}, \citenamefont
  {Sammak}, \citenamefont {Scappucci},\ and\ \citenamefont
  {Veldhorst}}]{nico2021}%
  \BibitemOpen
  \bibfield  {author} {\bibinfo {author} {\bibfnamefont {N.~W.}\ \bibnamefont
  {Hendrickx}}, \bibinfo {author} {\bibfnamefont {W.~I.~L.}\ \bibnamefont
  {Lawrie}}, \bibinfo {author} {\bibfnamefont {M.}~\bibnamefont {Russ}},
  \bibinfo {author} {\bibfnamefont {F.}~\bibnamefont {van Riggelen}}, \bibinfo
  {author} {\bibfnamefont {S.~L.}\ \bibnamefont {de~Snoo}}, \bibinfo {author}
  {\bibfnamefont {R.~N.}\ \bibnamefont {Schouten}}, \bibinfo {author}
  {\bibfnamefont {A.}~\bibnamefont {Sammak}}, \bibinfo {author} {\bibfnamefont
  {G.}~\bibnamefont {Scappucci}},\ and\ \bibinfo {author} {\bibfnamefont
  {M.}~\bibnamefont {Veldhorst}},\ }\bibfield  {title} {\bibinfo {title} {A
  four-qubit germanium quantum processor},\ }\href
  {https://doi.org/10.1038/s41586-021-03332-6} {\bibfield  {journal} {\bibinfo
  {journal} {Nature}\ }\textbf {\bibinfo {volume} {591}},\ \bibinfo {pages}
  {580} (\bibinfo {year} {2021})}\BibitemShut {NoStop}%
\end{thebibliography}%

\end{document}